\documentclass[aps,prb,twocolumn,superscriptaddress,showpacs]{revtex4-1}
\usepackage{graphicx,amsmath,color,placeins}
\usepackage{booktabs} %  引入三线表宏包
\usepackage{multirow} %  表格制作
\usepackage{setspace} %  使用间距宏包
\usepackage{lipsum} % 调用跨栏长公式宏包
\usepackage{epsfig,amssymb,subfigure,bm,dsfont}
\let\hide\iffalse
\usepackage[caption=false]{subfig}
\usepackage{epstopdf}
\usepackage{soul}
\usepackage{ulem}

\begin{document}

\title{The influence of high-energy local orbitals and electron-phonon interactions on the band gaps and optical absorption spectra of hexagonal boron nitride}
\author{Tong Shen}
\affiliation{State Key Laboratory for Artificial Microstructure and Mesoscopic Physics, Frontier Science Center for Nano-optoelectronics and School of Physics, Peking University, Beijing, China}
\author{Xiao-Wei Zhang}
\email{willzxw@pku.edu.cn}
\affiliation{International Center for Quantum Materials and School of Physics, Peking University, Beijing, China}
\author{Honghui Shang}
\affiliation{State Key Laboratory of Computer Architecture, Institute of Computing Technology, Chinese Academy of Sciences, Beijing, China}
\author{Min-Ye Zhang}
\affiliation{Beijing National Laboratory for Molecular Sciences, College of Chemistry and Molecular Engineering, Peking University, Beijing,China}
\author{Xinqiang~Wang}
\affiliation{State Key Laboratory for Artificial Microstructure and Mesoscopic Physics, Frontier Science Center for Nano-optoelectronics and School of Physics, Peking University, Beijing, China}
\affiliation{Collaborative Innovation Center of Quantum Matter, Peking University, Beijing 100871, P. R. China}
\author{En-Ge Wang}
\affiliation{International Center for Quantum Materials and School of Physics, Peking University,
Beijing, China}
\affiliation{Ceramic Division, Songshan Lake Lab, Institute of Physics, Chinese Academy of Sciences, Guangdong, China}
\affiliation{School of Physics, Liaoning University, Shenyang, China}
\author{Hong Jiang}
\email{jianghchem@pku.edu.cn}
\affiliation{Beijing National Laboratory for Molecular Sciences, College of Chemistry and Molecular Engineering, Peking University, Beijing, China}
\author{Xin-Zheng Li}
\email{xzli@pku.edu.cn}
\affiliation{State Key Laboratory for Artificial Microstructure and Mesoscopic Physics, Frontier Science Center for Nano-optoelectronics and School of Physics, Peking University, Beijing, China}
\affiliation{Collaborative Innovation Center of Quantum Matter, Peking University, Beijing, China}
\date{\today}
\begin{abstract}
We report \textit{ab initio} band diagram and optical absorption spectra of hexagonal
boron nitride ($h$-BN), focusing on unravelling how the completeness of basis set
for $GW$ calculations and how electron-phonon interactions (EPIs) impact on them.
The completeness of basis set, an issue which was seldom discussed in previous optical
spectra calculations of $h$-BN, is found crucial in providing converged quasiparticle band gaps.
In the comparison among three different codes, we demonstrate that by including high-energy local orbitals in the all-electron linearized augmented plane
waves based $GW$ calculations, the quasiparticle direct and fundamental indirect
band gaps are widened by $\sim$0.2~eV, giving values of 6.81 eV and 6.25 eV respectively
at the $GW_0$ level.
EPIs, on the other hand, reduce them to 6.62 eV and 6.03 eV respectively at 0 K, and 6.60 eV
and 5.98 eV respectively at 300 K.
With clamped crystal structure, the first peak of the absorption spectrum is at 6.07 eV, originating
from the direct exciton contributed by electron transitions around $K$ in the Brillouin zone.
After including the EPIs-renormalized quasiparticles in the Bethe-Salpeter equation, the
exciton-phonon coupling shifts the first peak to 5.83 eV at 300 K, lower than the experimental
value of $\sim$6.00 eV.
This accuracy is acceptable to an \textit{ab initio} description of excited states with no
fitting parameter.
\end{abstract}

\maketitle

\clearpage

\section{INTRODUCTION}
%
%\begin{itemize}
%\item wide bandgap
%\item high thermodynamical and chemical stability.
%\item graphene substate
%\item van der Waals heterostructures
%\item ultraviolet laser devices
%\end{itemize}
%
Hexagonal boron nitride ($h$-BN) is a wide band gap semiconductor with lamellar structure,
similar to graphite which is metallic.
Its wide band gap nature, low dielectric constant, and high thermal and chemical
stability mean that it is a promising candidate material in ultraviolet
optical and electronic devices~\cite{Geim2013,Dean2010,Watanabe2004,Schue2019}.
The $sp^{\text{2}}$-bonding results in two-dimensional (2D) honeycomb lattice
and atomically smooth surface.
As such, under the circumstances of the recently emerging and rapidly developing field of
layer engineering of 2D semiconductor conjunctions, monolayer or few-layer $h$-BN has
also been intensively employed for van der Waals heterostructures
construction~\cite{Geim2013,He2014,Chen2014,Jin2017}.
In spite of this popularity of $h$-BN, determination of the most basic property underlying its
applications in electronic and optical devices, i.e. its band diagram, remains unresolved.
High-resolution angle-resolved photoemission spectroscopy (ARPES) is the most direct method in
measuring the band diagram of crystals, and it has been applied to determine the position and dispersion
of its valence-band maxima~\cite{Henck2017}.
This information of its conduction-band states, however, has never been reported.
In fact, measuring conduction bands is a general technical challenge of ARPES for wide band gap
semiconductors~\cite{Damascelli2003}.
Therefore, one has to resort to optical experiments to answer this question in
an indirect manner.
In early optical measurements of photoluminescence (PL), the spectra were often interpreted to
indicate a direct band gap~\cite{Watanabe2004,Evans2008,Wanatabe2009}.
More recent experiments, on the other hand, present strong evidence of a indirect band gap
nature~\cite{Cassabois2016, Vuong2017a, Vuong2017b, Vuong2018, Schue2019}, as reviewed
in Ref.~\onlinecite{Caldwell2019}.
From the theoretical perspective, direct interpretation of the optical spectra is far from being
trivial~\cite{Hannewald2000, Pedro2016, Cannuccia2019, Paleari2019}.
Single particle excitation is the basis upon which all kinds of optical spectra were
calculated.
But as mentioned, accurate determination of the quasiparticle band diagram remains a question under
debate due to the absence of ARPES data of the conduction bands for comparison~\cite{Damascelli2003}.
Most often, one resorts to $GW$ approximation within the many-body perturbation
theory (MBPT)~\cite{Hybertsen1986, Godby1989}.
Or, alternatively, a scissor operator is applied directly to the Kohn-Sham eigenvalues.
Upon this, the excitonic effects and electron-phonon interactions (EPIs) can be
included~\cite{Marini2008}.
When analysis of the PL spectra is the target, apparatus like nonequilibrium Green's
function or density matrix theory will be resorted to~\cite{Pedro2016, Cannuccia2019, Paleari2019}.
Along this route, impressive progress has been achieved in understanding different optical spectra of
$h$-BN in recent years~\cite{Marini2008, Pedro2016, Paleari2018, Cannuccia2019, Paleari2019}.
But we note that some fundamental questions, which are crucial to a systematically
improvable \textit{ab initio} description of the electronic structures and optical properties
of $h$-BN, remain.
Most prominently, how does the description of quasiparticle band structure impact on the final
optical spectra is seldom discussed, in spite of the fact that it is the basis upon which all these
further analysis of the optical spectra is built.
Different codes may give different quasiparticle band
gaps\cite{Wirtz2005a,Arnaud2006,Paleari2018,Hunt2020}, and within $GW$ approximation
results obtained from different approaches like $G_0W_0$, $GW_0$, or self-consistent $GW$ also differ~\cite{Paleari2018,Paleari2019,Hunt2020}.
This is in sharp contrast with the situation of other well-known semiconductors, e.g. Si,
where theoretical descriptions of the optical spectra were obtained at a satisfactory and
most important systematically improvable basis~\cite{Ku2002,Delaney2004,Tiago2004,Friedrich2006,Shishkin2006,vanSchifgaarde2006,Gomez2008,Li2012,Jiang2016,Hinuma2014}.
Besides quasiparticle band diagram, discrepancies also exist on the influence of EPIs on the
renormalization of the quasiparticle and optical band gaps~\cite{Marini2008,Hunt2020,Mishra2019}.
In Ref.~\onlinecite{Marini2008}, it was reported that exciton-phonon couplings widen the direct
optical band gap of $h$-BN by 0.07 eV at 300 K.
While in Ref.~\onlinecite{Hunt2020}, EPIs were found to decrease the quasiparticle band gap by 0.4 eV
at the same temperature.
Considering the fundamental importance of this semiconductor in electronic and optical applications,
a thorough theoretical study on what the quasiparticle band diagram of $h$-BN is like and
how EPIs impact on the quasiparticle and optical properties is highly desired.
In this paper, we studied these problems from the theoretical perspective
using a combination of state-of-the-art \textit{ab initio} methods.
The completeness of the basis set for the $GW$ calculations, an issue which
had been often overlooked, is analyzed first.
This is done by using the benchmark electronic structure calculation method in solids, i.e.
the all-electron linearized augmented plane waves (LAPW) method.
Including high-energy local orbitals (HLOs) increases the direct and fundamental indirect
band gaps by $\sim$0.2 eV, when compared with previous \textit{ab initio} results.
Within $GW$ approximation, the $GW_0$ approach further widens these gaps by $\sim$0.3 eV when
compared with the standard $G_0W_0$ approach.
Our direct and fundamental $GW_0$ band gap with clamped crystal structure is 6.81 and 6.25 eV, respectively.
The absence of ARPES results means that a direct discrimination for the pros and cons of these
different numerical schemes is hard.
However, the fact that most \textit{ab initio} optical spectra must be blueshifted by $\sim$0.4-0.5 eV to
compare with experiments means that the $GW_0$ results obtained with well-converged basis
set outperform other choices in previous calculations for descriptions of the quasiparticle band
diagram~\cite{Pedro2016, Paleari2018, Cannuccia2019, Paleari2019}.
Based on these $GW$ calculations with clamped crystal structure, the excitonic effects and the
influence of EPIs on the quasiparticle and optical gaps were analyzed.
The direct (fundamental indirect) band gap is reduced by $\sim$0.2 eV to 6.62 (6.03) eV at 0 K,
and 6.60 (5.98) eV at 300 K.
Using the Bethe-Salpeter equation (BSE), the excitonic effects were first calculated using clamped
crystal structure and we found an exciton binding energy of 0.76 eV, associated with the direct transitions
around $K$ in the Brillouin zone (BZ).
The quasiparticle renormalizations due to EPIs were investigated within the MBPT and then were
incorporated into the BSE to study the exciton-phonon couplings.
The final influence of EPIs on the absorption spectra is a redshift of the peak position, which
results from a competition between the decrease of the quasiparticle band gaps which shifts the
absorption peak to lower energy, and the decrease of the exciton binding energy which shifts it
to the higher energy.
The former factor is dominant, meaning that EPIs redshift both the quasiparticle band gaps and the
optical absorption peak by $\sim$0.2 eV.
This is qualitatively different from Ref.~\onlinecite{Marini2008}, where the EPIs blueshift the
absorption peak by 0.07 eV.
The magnitude for the redshift of quasiparticle band gaps is also smaller than the value
of 0.4 eV in Ref.~\onlinecite{Hunt2020}.
Our final absorption line shape agrees well with available experiments, with the position of the
absorption peak redshifted by $\sim$0.2 eV.
With issues like the starting point of optical spectra calculations (i.e. quasiparticle band diagram)
and the influence of EPIs on quasiparticle/optical band gaps clarified, this is
acceptable for \textit{ab initio} calculations.
Our manuscript is organized as follows.
In Sec. II we present a short overview on the implementation of the $GW$ in the LAPW basis and a
clear account of the methodology for \textit{ab initio} calculations of EPIs.
The computational details of our calculations were also provided.
In Sec. III, we report the influence of including HLOs in LAPW-based $GW$ calculations.
The $T$-dependent band renormalization and the zero- and finite-$T$ optical absorption incorporating both excitonic effect and EPIs of $h$-BN were also analyzed.
The conclusions are given in Sec. IV.

\section{Methods and computational details}
\subsection{$GW$ in the LAPW basis including HLOs}
The LAPW method is the benchmark method for performing electronic structure calculations
of crystals~\cite{Wien2k}.
It stems from the augmented plane waves (APW) method, where the muffin-tin (MT) approximation
originally proposed by J. C. Slater is used~\cite{Slater1937}:
the wave functions are atomic-like in the region close to nuclei defined by the MT
radius $R_{\rm MT}$, and plane-wave-like in the interstitial region ($\bm I$) between nuclei.
Inside the atomic spheres, atomic-like wave function is a linear combination of radial function
times spherical harmonics function.
Here, $\mathbf{r}^{\alpha} \equiv \mathbf{r}-\mathbf{r}_{\alpha}$, $R_{\rm MT}^{\alpha}$ is the MT radius of the $\alpha$th atom centered at $\mathbf{r}_\alpha$.
$u_{\alpha l}(r^\alpha;E_{\alpha l})$ is the solution of the radial Schr\"{o}dinger equation at a
fixed reference energy $E_{\alpha l}$ in the spherical potential of the respective MT sphere,
and $\dot u_{\alpha l}(r^\alpha;E_{\alpha l})$ is its energy derivative.
The expansion coefficients include $A_{\alpha lm}(\mathbf k+\mathbf G)$ and
$B_{\alpha lm}(\mathbf k+\mathbf G)$, which are determined from the continuity of the basis
functions and their first derivative at the boundary of the MT sphere.
The LAPW basis can be reduced to an APW basis when $B_{\alpha lm}(\mathbf k+\mathbf G)=0$.
To make the LAPW basis set complete for the expansion of the electronic wave functions,
local orbitals (LOs) can be supplemented, as in Eq.~[\ref{LO}]~\cite{Jiang2016}
\begin{widetext}
\begin{equation}
\phi_{\rm \mathbf G}^{\rm \mathbf k}(\mathbf r)=
    \begin{cases}
    \sum_{lm}[A_{\alpha lm}(\mathbf k+\mathbf G)u_{\alpha l}(r^\alpha;E_{\alpha l})
    +B_{\alpha lm}(\mathbf k+\mathbf G)\dot u_{\alpha l}(r^\alpha;E_{\alpha l})]
    Y_{lm}(\hat{\mathbf r}^{\alpha}),
    & r^{\alpha} < R_{MT}^{\alpha}.\\
    \frac{1}{\sqrt\Omega}e^{i(\rm \mathbf k + \rm \mathbf G)\cdot \mathbf r}, & \mathbf r \in \mathbf \bm I.
    \end{cases}
\label{LAPW}
\end{equation}

\begin{equation}
\phi_{lm}^{\rm LO}(\mathbf r)=
    \begin{cases}
    [A_{\alpha lm}^{\rm LO}u_{\alpha l}(r^\alpha;E_{\alpha l})
    +B_{\alpha lm}^{\rm LO}\dot u_{\alpha l}(r^\alpha;E_{\alpha l})
    +C_{\alpha lm}^{\rm LO}u_{\alpha l}(r^\alpha;E_{\alpha l}^{(2)})]
    Y_{lm}(\hat{\mathbf r}^{\alpha}),
    &r^{\alpha}<R_{MT}^{\alpha}.\\
    0, & \mathbf r \in \mathbf \bm I.
    \end{cases}
\label{LO}
\end{equation}
\end{widetext}
with $E_{\alpha l}^{(2)}$ chosen to be close to the energy of the low-lying or high-lying states
of interest and the coefficients $A_{\alpha lm}^{\rm LO}$, $B_{\alpha lm}^{\rm LO}$,
$C_{\alpha lm}^{\rm LO}$ are determined by the requirement that $\phi_{lm}^{\rm LO}(\mathbf r)$
is normalized and is zero in value and slope at the MT sphere boundary.
Complement to the conventional understanding that LOs are important for a proper description
of quasiparticle energies in system with semicore states~\cite{Rohlfing1998,Li2012},
recently, it was realized that the LOs of high-lying unoccupied states are also crucial
in getting the quasiparticle band gaps converged~\cite{Jiang2016,Nabok2016}.
The parameter $n_{\rm LO}$ is the additional number of nodes of the high-lying LOs with respect to the corresponding valence orbital and $l_{\rm max}^{\rm LO}$ is the maximal angular quantum number of LOs, which represent the accuracy of the LO-enhanced LAPW basis.
Without these high-lying LOs, the basis set cannot be complete for the description of the excited
state wave functions~\cite{Krasovskii1994}, no matter how many LAPWs are used for the expansion
of the wave functions~\cite{Jiang2016, Nabok2016}.
This is analogous to the case when quasiparticle energies are calculated using pseudopotential (PP)
based methods.
Conventional PPs for density-functional theory (DFT) calculations cannot guarantee well-converged
$GW$ band gaps, no matter how many plane waves are used~\cite{Tiago2004, Shishkin2006}.
One needs additional channels to be included in the construction of the new PPs for the $GW$
calculations to be performed.
And these channels are designed to describe the high-lying electronic states.
In Sec. III, we will give a detailed discussion on how this issue will impact on the convergence
of the $h$-BN quasiparticle band gaps.

\subsection{Electron-phonon interaction}
The standard form of the Hamiltonian in describing an electron-phonon coupled system using second-quantized formalism can be written as~\cite{Giustino2017}
\begin{equation}
\begin{split}
\hat{H}
=&\hat H_{\text{e}}+\hat H_{\text{ph}}+\hat H_{\text{e-ph}} \\
=&\sum_{n\mathbf{k}}\varepsilon_{n\mathbf{k}}\hat c^{\dagger}_{n\mathbf k}\hat c_{n\mathbf k}
+\sum_{\mathbf{q}\nu}\hbar\omega_{\mathbf{q}\nu}(\hat{a}^{\dagger}_{\mathbf{q}\nu}\hat{a}_{\mathbf{q}\nu}+1/2)
\\
&+N^{-1/2}_{p}\sum_{\substack{\mathbf{k,q}\\nn^{'}\nu}}g^{\rm Fan}_{nn^{'}\nu}(\mathbf{k,q})\hat c^{\dagger}_{n\mathbf{k+q}}\hat c_{n^{'}\mathbf k}(\hat a_{\mathbf{q}\nu}+\hat a^{\dagger}_{-\mathbf{q}\nu})
\\
&+N^{-1}_{p}\sum_{\substack{\mathbf{k,q,q^{'}}\\nn^{'}\nu^{'}}}g^{\rm DW}_{nn^{'}\nu\nu^{'}}(\mathbf{k,q,q^{'}})\hat c^{\dagger}_{n\mathbf{k+q+q^{'}}}\hat c_{n\mathbf k}\\
&\times(\hat a_{\mathbf{q}\nu}+\hat a^{\dagger}_{-\mathbf{q}\nu})(\hat a_{\mathbf{q^{'}}\nu^{'}}+\hat a^{\dagger}_{-\mathbf{q^{'}}\nu^{'}}).
\label{H}
\end{split}
\end{equation}
In this expression
$\hat H_{\text{e}}=\sum_{n\mathbf{k}}\varepsilon_{n\mathbf{k}}\hat c^{\dagger}_{n\mathbf k}\hat c_{n\mathbf k}$ is the Hamiltonian of the electron subsystem with the $\varepsilon_{n\mathbf{k}}$ being the eigenvalue ($n$ and $\bf{k}$ are band and crystal momentum index, respectively).
$\hat{c}_{n\bf{k}} (\hat{c}^{\dagger}_{n\bf{k}})$ is the electron annihilation (creation) operator.
$\hat{H}_{\text{ph}}=\sum_{\mathbf{q}\nu}\hbar\omega_{\mathbf{q}\nu}(\hat{a}^{\dagger}_{\mathbf{q}\nu}\hat{a}_{\mathbf{q}\nu}+1/2)$ is the Hamiltonian of the phonon subsystem with the $\omega_{\mathbf{q}\nu}$ being the phonon frequency ($\nu$ and $\bf{q}$ are branch and crystal momentum index, respectively).
$\hat{a}_{\bf{q}\nu} (\hat{a}^{\dagger}_{\bf{q}\nu})$ is the phonon annihilation (creation) operator.
Getting these two terms belongs to the single-particle problem in Eq.~[\ref{H}], and in practical
calculations they are often described using the Kohn-Sham orbitals and the phonon obtained from DFT.
The electron-phonon coupling Hamiltonian $\hat{H}_{\text{e-ph}}$ consist of two terms
$\hat{H}_{\text{e-ph}}^{(1)}$ and $\hat{H}_{\text{e-ph}}^{(2)}$.
$\hat{H}_{\text{e-ph}}^{(1)}$ is the first order coupling in terms of the atomic
displacement (the third line) and $\hat{H}_{\text{e-ph}}^{(2)}$ is the second order coupling in terms of
the atomic displacement (the last two lines).
The corresponding electron-phonon coupling matrix elements are $g^{\rm Fan}_{nn^{'}\nu}$ and $g^{\rm DW}_{nn^{'}\nu\nu^{'}}$ as defined below.
$N_{p}$ is the number of the unit cells in the supercell associated with the transformation
between the real-space and normal-mode coordinates.
Starting from the coupled electron-phonon Hamiltonian in Eq.~[\ref{H}],	
one can obtain the electron-phonon self-energy using the MBPT, which renormalizes the
quasiparticle energies.
Due to the complexity of higher-order self-energy, one often keeps only two terms up
to the second order, i.e. the Fan self-energy (the second-order self-energy of $\hat{H}_{\text{e-ph}}^{(1)}$) and the Debye-Waller (DW) self-energy (the first-order self-energy of $\hat{H}_{\text{e-ph}}^{(2)}$).
Specifically,
\begin{equation}
\begin{split}
\Sigma_{n\mathbf{k}}^{\rm Fan}(\omega,T)=&\sum_{n^{\prime}\mathbf{q}\nu}
\frac{\big|g_{nn^{\prime}\nu}^{\rm{Fan}}(\bf{k},\bf{q})\big|^2}{N}\\
&\left[\frac{{n_{\mathbf{q}\nu}(T)}+1-f_{n^{\prime}\mathbf{k-q}}}
{\omega-\varepsilon_{n^{\prime}\mathbf{k-q}}-\omega_{\mathbf{q}\nu}+i\eta \text{sgn}(\omega)}\right.\\
&\left.+\frac{n_{\mathbf{q}\nu}(T)+f_{n^{\prime}\mathbf{k-q}}}
{\omega-\varepsilon_{n^{\prime}\mathbf{k-q}}-\omega_{\mathbf{q}\nu}+i\eta \text{sgn}(\omega)}\right],
\end{split}
\label{Fan}
\end{equation}
\begin{equation}
\Sigma_{n\mathbf{k}}^{\rm DW}(T)=-\frac{1}{2}\sum_{n^{\prime}\mathbf{q}\nu}
\frac{g_{nn^{\prime}\nu}^{\rm DW}(\bf{k},\bf{q})}{N}
\left[\frac{2n_{\mathbf{q}\nu}(T)+1}
{\varepsilon_{n\mathbf{k}}-\varepsilon_{n^{\prime}\mathbf{k}}}\right].
\label{DW}
\end{equation}
Here, $n_{\mathbf{q}\nu}$ and $f_{n^{\prime}\mathbf{k-q}}$ correspond to the Bose-Einstein and Fermi-Dirac distribution functions, while $N$ is the number of $\mathbf{q}$ points in the BZ.
$g_{nn^{\prime}\nu}^{\rm Fan}(\bf{k},\bf{q})$ is the first-order gradient of the Kohn-Sham
self-consistent potential $v^{\text{KS}}$ with respect to the atomic displacement,
\begin{equation}
g_{nn^{\prime}\nu}^{\rm Fan}(\bf{k},\bf{q}) = \left<\textit{u}_{n\rm\mathbf{k+q}}|\Delta_{\rm\mathbf{q}\nu}\textit{v}^{\rm{KS}}|\textit{u}_{n^{\prime}\rm\mathbf{k}}\right>_{\rm{uc}},
\label{EPC}
\end{equation}
with $u_{n\rm\mathbf{k}}$ and $u_{n\rm\mathbf{k+q}}$ being the Bl\"och-periodic components of the KS electron wave functions.
``uc'' means that the integral is performed within the unit cell.
$g_{nn^{\prime}\nu\nu^{\prime}}^{\rm DW}(\bf{k},\bf{q}^{\prime},\bf{q})$ is the second-order gradient of the Kohn-Sham potential with respect to the atomic displacement
\begin{equation}
g_{nn^{\prime}\nu\nu^{\prime}}^{\rm DW} (\bf{k}, \bf{q}, \bf{q}^{\prime})= \frac{1}{2} \left<\textit{u}_{n\rm\mathbf{k+q+q^{\prime}}}|\Delta_{\rm \mathbf{q}\nu}\Delta_{\rm\mathbf{q^{\prime}}\nu^{\prime}}\textit{v}^{\rm{KS}}|\textit{u}_{n^{\prime}\rm\mathbf{k}}\right>_{\rm{uc}}.
\label{EPC}
\end{equation}
Note that we have used the translational invariance to relate Eq.~[\ref{EPC}] to the first-order gradient of the Kohn-Sham potential and obtain Eq.~[\ref{DW}]{~\cite{Ponce2014, Ponce2014b}}.
In the quasiparticle approximation (QPA), the $T$-dependent quasiparticle energies can be written as
\begin{equation}
E_{n\mathbf{k}}(T)=\varepsilon_{n\mathbf{k}}+Z_{n\mathbf{k}}(T)\left[
\Sigma_{n\mathbf{k}}^{\rm Fan}
(\varepsilon_{n\mathbf{k}},T)+\Sigma_{n\mathbf{k}}^{\rm DW}(T)\right],
\label{band}
\end{equation}
where $Z_{n\mathbf{k}}(T)=\left[1-\frac{\partial{\text{Re}\Sigma_{n\mathbf{k}}^{\rm Fan}(\omega)}}
{\partial{\omega}}\big|_{\omega=\varepsilon_{n}}\right]^{-1}$ is the renormalization factor.
The Fan self-energy is frequency-dependent while the DW self-energy is frequency-independent.
Thus the former not only shifts the electronic state energy but also give finite quasiparticle
lifetime, and the latter only shifts the energy.

\subsection{Computational details}
In this work, experimental lattice parameters ($a$ = 2.504 and $c$ = 6.661 \AA) are used to
perform all the calculations~\cite{Solozhenko1995}.
Discussions on the van der Waals interactions can be found in Supplemental Materials.
This is consistent with the routine in previous theoretical simulations of the
quasiparticle energies and optical spectra~\cite{Ku2002,Delaney2004,Tiago2004,Friedrich2006,Shishkin2006,vanSchifgaarde2006,Arnaud2006,Marini2008,Gomez2008,Li2012,Jiang2016}.
To compare computational results obtained using different numerical implementations of DFT and the $GW$
approximation method, three kinds of codes were chosen, including: (i) the all-electron linearized augmented plane waves method based WIEN2k~\cite{Schwarz2002} and GAP2~\cite{Jiang2013,Jiang2016}, (ii) the projector-augmented-wave (PAW) method based Vienna \textit{Ab initio} Simulation
Package (VASP)~\cite{Kresse1996}, and (iii) the pseudopotentials method based QUANTUM ESPRESSO (QE)~\cite{Giannozzi2009} and YAMBO~\cite{Marini2009,Sangalli2019}.
Local-density approximation (LDA) is chosen for the DFT calculations.
In the WIEN2k calculations, LDA results presented in Table~\ref{table1} are obtained with
a $12\times12\times4$ k-mesh, and we have chosen $R_{\text{MT}}$(B,N) = (1.23, 1.35) Bohr and $RK_{\text{max}}$ = 7.0.
In the VASP calculations, the PAW potentials were used along with an 800~eV plane wave cutoff
energy and the band gaps were converged with a $6\times6\times2$ k-point grid.
In the QE calculations, the DFT ground state were obtained using LDA norm-conserving pseudopotentials
with a kinetic energy cut-off at 110 Ry and a $6\times6\times2$ k-point grid.
Quasiparticle band structures are calculated by using the $GW$ approximation.
Two approaches, i.e. $G_0W_0$ and $GW_0$, were used.
With $G_0W_0$, the one-body Green's function and the screened Coulomb interaction were calculated
directly from the LDA Kohn-Sham orbitals, and the self-energy was obtained from them in the one-shot
manner~\cite{Hybertsen1986}.
With $GW_0$, a self-consistent treatment was applied to the Green's function by updating the
quasiparticle energies~\cite{Shishkin2007, Jiang2010}.
The screened Coulomb interaction keeps its form as in $G_0W_0$.
The all-electron LAPW based $G_0W_0$ and $GW_0$ calculations were performed
using the GAP2 code~\cite{Jiang2013}.
The results are converged at $12\times12\times4$ k-point grid, $n_{\rm LO}=2$, and $l_{\rm max}^{\rm LO}=4$.
The details are explained in Fig. S4.
In the VASP calculations, PAW pseudopotentials were used along with a 800~eV
plane wave cutoff energy and the band gaps were converged with a $12\times12\times4$ k-point grid.
For the frequency dependence of the screened Coulomb interaction, different treatments are used in the three codes that we have used here. In the GAP2 code, the correlation self-energy was first calculated along the imaginary axis and then analytically continued into the real frequency axis by using the multi-pole fitting scheme, as detailed in Ref. \onlinecite{Jiang2013}. In the VASP code, the full-frequency calculation on the real axis was used~\cite{Shishkin2006}. In the YAMBO code, the Godby-Needs generalized plasmon-pole model was used for the $GW$ self-energy~\cite{Godby1989}.
The optical absorption spectra including excitonic effects were obtained using YAMBO, where we have employed
a scissor operator of 2.314 eV to correct the eigenvalues of the 6th to 10th bands in
solving the BSE.
This scissor is obtained from \textit{ab initio} calculations, by taking the difference between
the $GW_0$ and LDA direct band gaps.
A dense uniform $18\times18\times6$ k-point grid is necessary to converge the optical absorption
spectra.
Electron-phonon interactions were considered by combining QE and YAMBO.
The phonon dispersion was obtained using a uniform $12\times 12\times 8$ sampling of the electron BZ and a uniform $8\times 8 \times 6$ sampling of the phonon BZ.
Electron-phonon self-energies were obtained using 600 random $\mathbf{q}$ points in the phonon BZ, a uniform $6\times6\times2$ k-grid for the electron BZ and 300 electronic bands.
The convergence tests are given in Supplemental Materials.
The QPA was used to correct the LDA eigenvalues based on electron-phonon self-energies.
In so doing, the real parts of quasiparticle energies give $T$-dependent band gaps and the imaginary parts give the life time due to electron-phonon renormalizations.
Finally, these quasiparticle energies were thrown into the $T$-dependent BSE and exciton-phonon coupling was considered this way~\cite{Marini2008}.

\section{Results and Discussion}
\subsection{Quasiparticle band structure}
\begin{figure}[]
\includegraphics[width=0.93\linewidth]{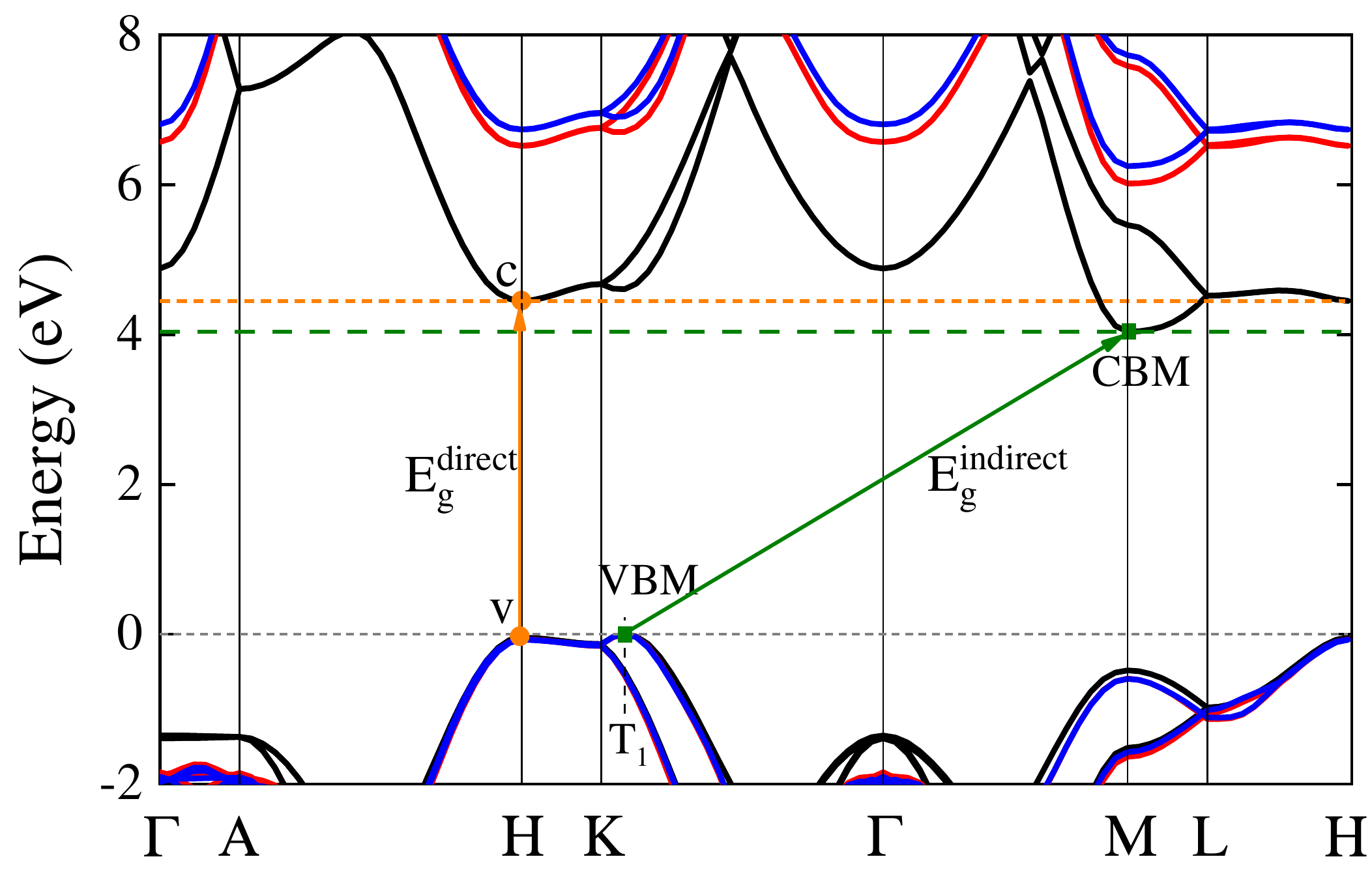}
\caption{\label{FIG1}  Electronic band structure calculation by means of LAPW method shows LDA (black solid lines) and $GW_0$ without/with HLOs (red/blue solid lines) results of $h$-BN. The orange arrow line represents direct gap, and the green one represents fundamental gap.}
\end{figure}

Since the quasiparticle band structure is the basis upon which the electron-hole interactions and
EPIs are analyzed, we start by looking at the accuracy and convergence of the band diagrams, in the theoretical calculations.
Using the standard LAPW basis set, the LDA and $GW$ band diagram obtained from Wien2k and GAP2 were
shown by black and red solid lines in Fig.~\ref{FIG1}.
The smallest direct band gap is at $H$ [$\mathbf{k}=(-1/3,2/3,1/2)$] from $v$ to $c$.
The fundamental gap, however, is an indirect one between $T_1$ (close to $K$ [$\mathbf{k}=(-1/3,2/3,0)$])
and $M$ [$\mathbf{k}=(0,1/2,0)$] (denoted by VBM and CBM).
The LDA fundamental and direct band gaps are 4.04 and 4.50 eV, respectively.
The $GW_0$ with the standard LAPW gives fundamental and direct band gaps of 6.02 and 6.59 eV, respectively, and the band diagram of $GW_0$ is basically a rigid shift from the LDA one.
These results are consistent with previous reports.
To demonstrate this consistency and unravel some details underlying previous optical spectra in
a clear manner, we compare the LDA, $G_0W_0$, and $GW_0$ band gaps obtained from the standard
treatment in Wien2k+GAP2, in VASP and in QE+YAMBO in Table~\ref{table1}.
The computational setup was given in Sec. II.C.
The $GW$ results of the YAMBO code come from Refs.~\onlinecite{Paleari2018, Paleari2019}.
Here by ``standard'', we mean the fact that no special care was taken to the descriptions of the
high-lying electronic states.
In other words, the basis set is complete for DFT calculations, when occupied states are the
determinant issue for the convergence of the computational outcome.
However, this completeness could be questionable for descriptions of the excited-state properties,
when high-lying electronic states are also
important~\cite{Tiago2004,Friedrich2006,Shishkin2006,vanSchifgaarde2006,Arnaud2006,Marini2008,Gomez2008,Li2012,Jiang2016}.
In LAPW-based calculations, this means that no additional HLOs were used~\cite{Jiang2016, Nabok2016}.
In PP-based calculations, this means that no additional channels were added in constructing
the PPs~\cite{Shishkin2006}.
Without these HLOs or additional channels, the basis set cannot be converged in the excited-state
electronic structure calculations solely by adding the number of plane waves.
In all three kinds of codes, $GW_0$ widens the band gaps by $\sim$0.3 eV when compared to $G_0W_0$.
Considering the fact that theoretical optical peaks are systematically lower than
experimental observations, this explains why $GW_0$ stands for a better starting point than
$G_0W_0$ in previous theoretical
simulations~\cite{Marini2008, Pedro2016, Paleari2018, Cannuccia2019, Paleari2019}.
Beside this, another important message one can get from this consistency between
results from different kinds of codes is that numerical errors from issues like
frequency integrations for the self-energy and construction of PPs are under control.
This is especially true when the VASP band gaps are compared with the Wien2k+GAP2 ones.
Considering the all-electron feature of the PAW pseudopotentials, this is not
surprising~\cite{Lejaeghere2016}.
Even with norm-conserving PPs, the direct and fundamental $GW_0$ band gaps agree within
0.06 eV and 0.13 eV with the all-electron LAPW-based ones.
Then, motivated by recent separate studies of Jiang \textit{et al.}\cite{Jiang2016} and
Nabok \textit{et al.}~\cite{Nabok2016} on the role played by HLOs in $GW$
calculations, we investigate how the $G_0W_0$ and $GW_0$ band gaps of $h$-BN can be
changed by including HLOs in the all-electron LAPW-based $GW$ calculations.
The results are shown by blue solid lines in Fig.~\ref{FIG1}, compared with the red ones without HLOs.
It is clear that the inclusion of HLOs has a significant influence on the $GW$ corrections,
which enlarges the band gaps by $\sim$0.2~eV, for both the direct and the fundamental ones.
This enlargement, seemingly small, stands for an important progress toward a systematically
improvable \textit{ab initio} description of the optical spectra of $h$-BN, considering
the fact that most previous \textit{ab initio} optical spectra must be blueshifted by $\sim$0.4-0.5 eV
to compare with experiments~\cite{Pedro2016, Paleari2018, Cannuccia2019, Paleari2019}.

\begin{figure}[]
\includegraphics[width=0.90\linewidth]{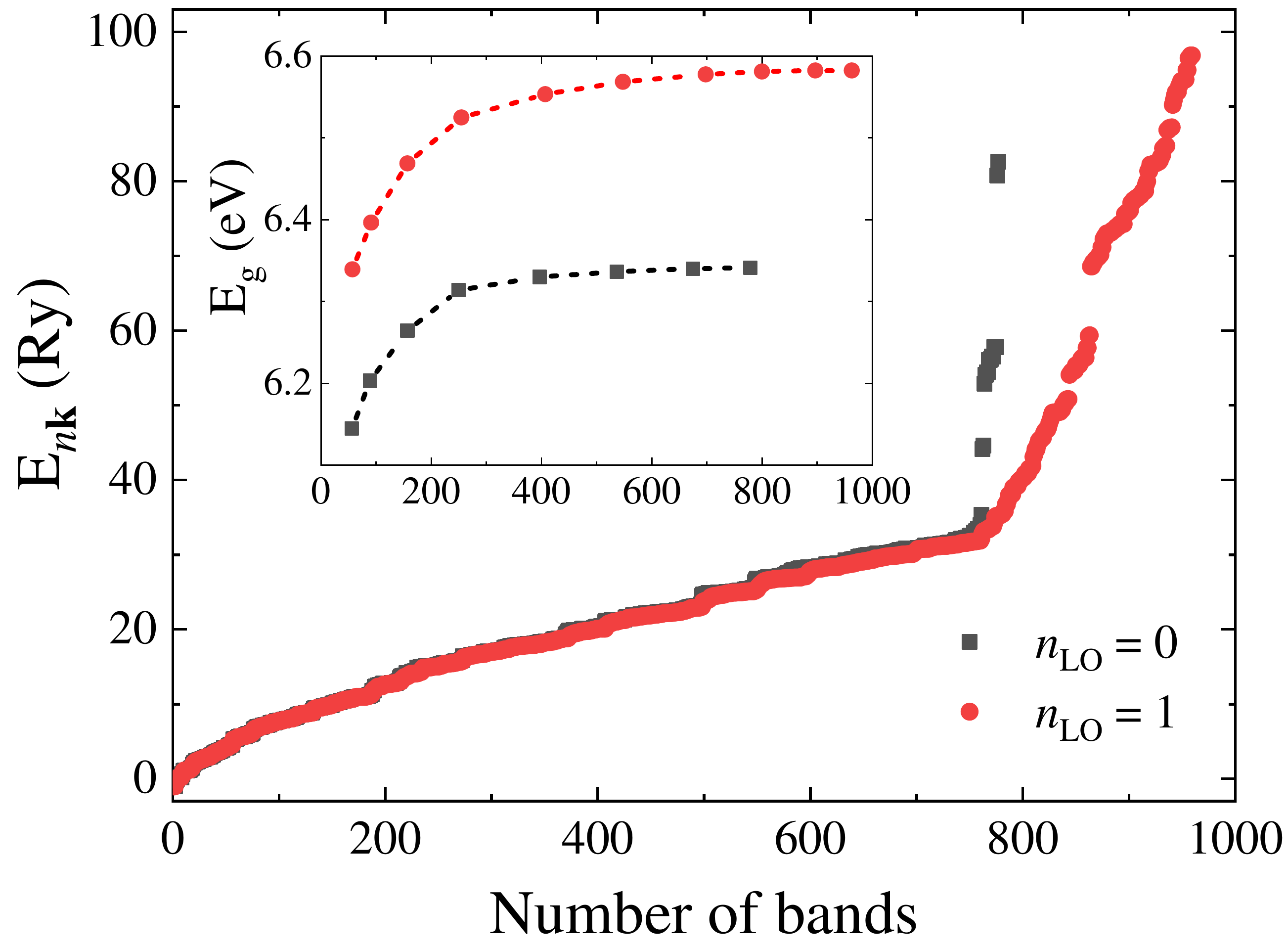}
\caption{\label{FIG2} Comparison of KS band energies of $h$-BN at the $\Gamma$ point obtained from using the default LAPW basis and those with additional high-energy local orbitals with $n_{\rm{LO}} =1$ and $l_{\rm max}^{\rm LO} = 4$. The inset shows the convergence of the $G_0W_0$ band gap of $h$-BN (calculated with $N_k = 2\times2\times1$) as a function of the number of bands considered with $n_{\rm{LO}} =0$ and $n_{\rm{LO}} =1$, respectively.
}
\end{figure}

For an in-depth understanding of this impact of HLOs on $GW$ band gap of $h$-BN,
we show the influence of additional HLOs on the KS single-particle spectrum.
Fig.~\ref{FIG2} gives the band energies of $h$-BN at the $\Gamma$ point [$\mathbf{k}=(0,0,0)$] obtained with $n_{\rm LO}$ = 0 (the default LAPW basis) and $n_{\rm LO}$ = 1 (including the HLOs in the LAPW basis), respectively.
Both of these two sets of data show that the energy of unoccupied states increases
smoothly as a function of the band index (nbands) up to the plane-wave cutoff
energy $\varepsilon_{\rm max}^{\rm PW}$ (36.0 Ry in the current case).
Then, it increases rapidly to much higher energy with the
data of $n_{\rm LO}$ = 0 increasing faster than $n_{\rm LO}$ = 1, meaning that adding LOs tends
to decrease the energies of these states.
It is worth noting that adding additional HLOs can improve the description
of high-lying unoccupied states and make additional high-energy states available.
In the inset of Fig.~\ref{FIG2}, we show the convergence of the $G_0W_0$ direct band gap as
a function of the total number of states (nbands).
Without HLOs, the band gap converges, but to the wrong value.
Therefore, both the accuracy of the conduction-band states within the plane-wave cutoff
and the availability of the high-energy LO states beyond the plane-wave cutoff are important
for the numerically accurate $G_0W_0$ band gap.
Specifically, including HLOs in LAPW-based $GW$ results widens the direct (fundamental) band
gap of $h$-BN by 0.22 (0.23) eV (Table~\ref{table1}).
Considering the fact that the quasiparticle excitation is the basis upon which the absorption
spectrum is calculated, a built-in underestimation of the absorption energy should exist at
the quasiparticle level when this issue related to the completeness of the basis set is neglected.
This might be the case in many previous theoretical calculations of the optical spectra.

\renewcommand\arraystretch{2}
\begin{table}[]
\setlength{\abovecaptionskip}{0.cm}
\setlength{\belowcaptionskip}{-0.cm}
\scriptsize
\centering
\caption{Calculated band gaps (in eV) from different theoretical approaches of $h$-BN. The influence
of HLOs on the $GW$ band gaps is highlighted.}
\label{table1}
\begin{tabular*}{\hsize}{@{}@{\extracolsep{\fill}}cccccccc@{}}
\toprule\\
\hline\hline
\qquad & \multicolumn{3}{c}{without HLOs} & \multicolumn{2}{c}{HLOs} \\
\hline
     WIEN2k + GAP2
     &LDA &$G_0W_0$ &$GW_0$
     &$G_0W_0$ &$GW_0$ \\
fundamental &4.04 &5.71 &6.02 &5.90 &6.25 \\
Direct   &4.50 &6.26 &6.59 &6.45 &6.81 \\
\hline
     VASP
     &LDA &$G_0W_0$ &$GW_0$
     &$G_0W_0$ &$GW_0$ \\
fundamental &4.04 &5.73 &6.03 &/    &/  \\
Direct   &4.51 &6.29 &6.61 &/    &/  \\
\hline
     QE + YAMBO
     &LDA &$G_0W_0$ &$GW_0$
     &$G_0W_0$ &$GW_0$  \\
fundamental &4.06 &5.73 &5.96 &/    &/  \\
Direct   &4.51 &6.24 &6.46 &/    &/  \\
\hline\hline
\bottomrule
\end{tabular*}
\end{table}
\subsection{Band gap renormalization}
\label{bandgaprenorm}
\begin{figure}[]
\includegraphics[width=0.83\linewidth]{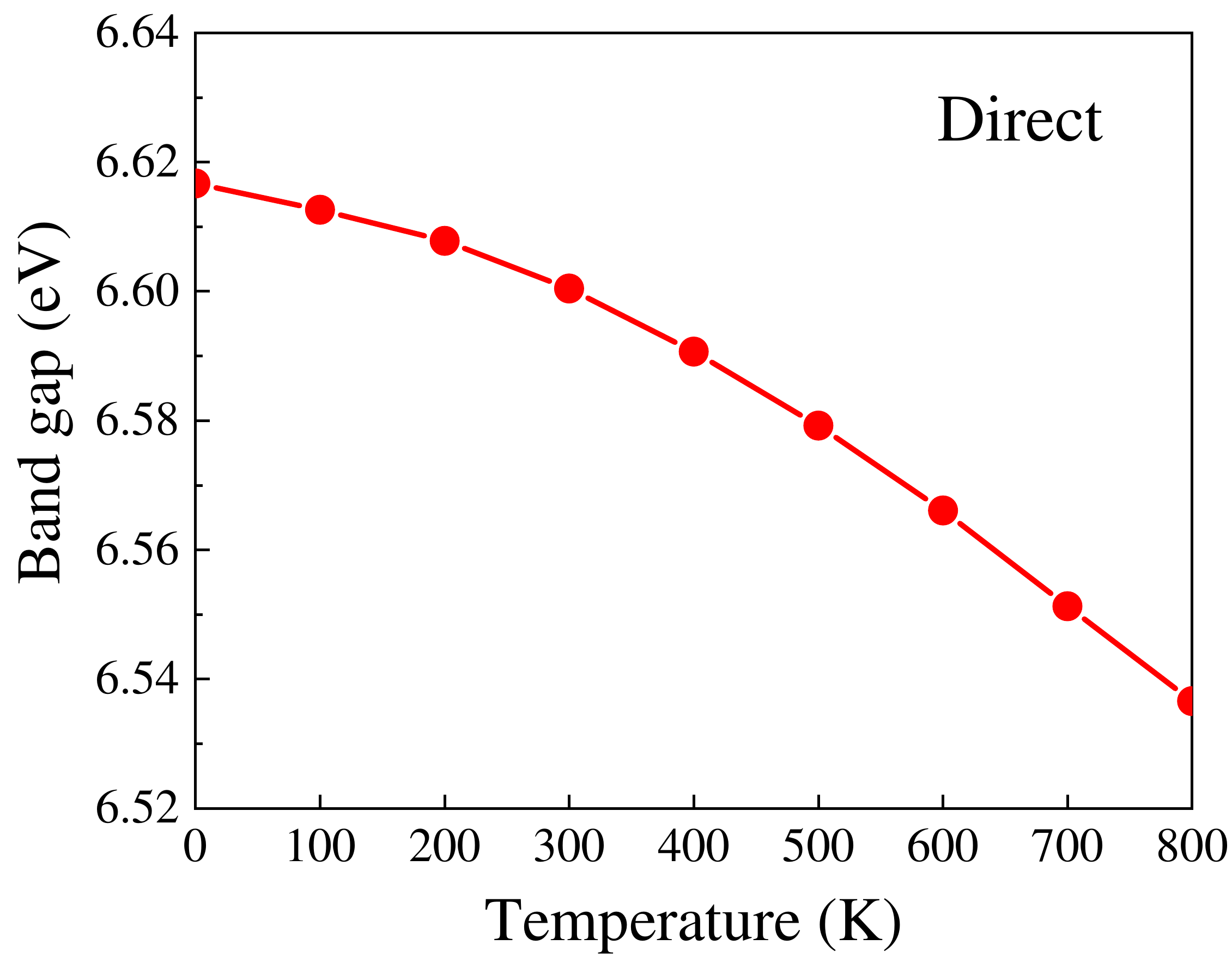}
\caption{\label{ZPR-2} $T$-dependent direct band gap of $h$-BN calculated using dynamical HAC theory. The band gap renormalization is added to the LAPW+HLOs corrected $GW_0$ gap, obtained in Sec. III.A, which is 6.82 eV in YAMBO code.}
\end{figure}

The above calculations have adopted a clamped crystal structure.
However, even at 0 K, the nuclei are not fixed and they vibrate around the equilibrium positions.
This motion is known as zero-point motion and its influence on the quasiparticle excitation energy
is known as zero-point renormalization to the quasiparticle excitation energy.
At finite $T$s, the EPIs also influence the quasiparticle excitation energy, with contributions from
both classical and quantum motions of the nuclei.
At high $T$s, the quantum contribution to the motion of the nuclei is negligible and the electron and
phonon interacts through a classical manner.
Such $T$-dependent renormalization of the single particle excitation energy due to EPIs can be
calculated using Eq.~[\ref{band}], and so can the fundamental and direct band gaps.
In Fig.~\ref{ZPR-2}, we show the EPIs-induced correction to the direct band gap of
$h$-BN (obtained using clamped crystal structure).
%\textcolor{blue}{The black dashed line is the result obtained by ....}
%
%It corresponds to the case when the finite-$T$ nuclei motion is described using classical mechanics.
%
The red circles are results obtained using Eq.~[\ref{band}], when both classical and
quantum contributions to the nuclear motion are taken into account.
It is clear that the EPIs-induced correction to the static direct band
gap has a weak $T$-dependence below room $T$ (300 K), and increases almost linearly beyond it.
The zero-point renormalization to the direct band gap is 0.20 eV, larger than many
typically semiconductor, which means that the EPIs are strong in $h$-BN.
\begin{figure}[]
\includegraphics[width=0.83\linewidth]{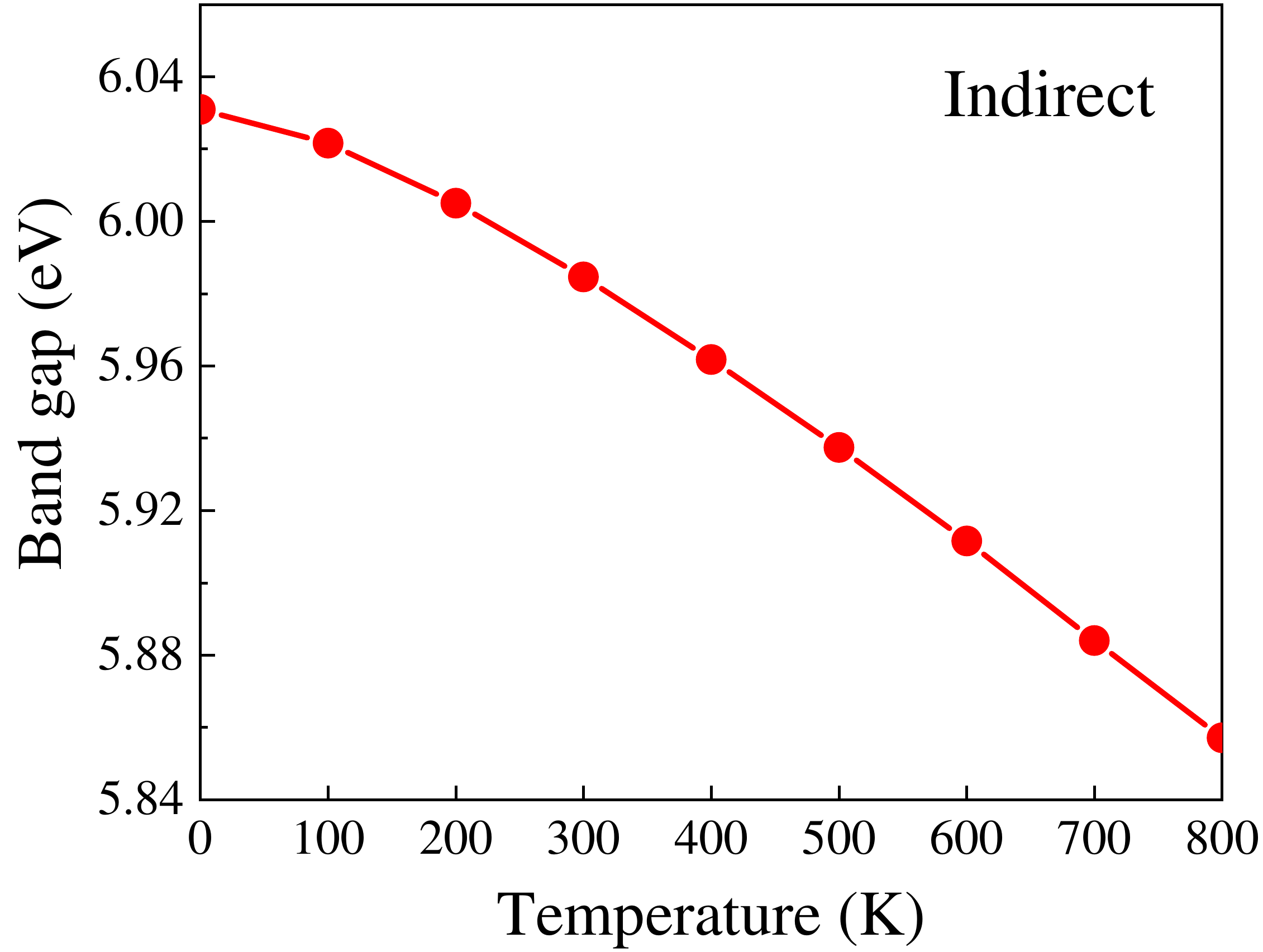}
\caption{\label{ZPR}  $T$-dependent fundamental band gap of $h$-BN calculated using dynamical HAC theory. The band gap renormalization is added to the LAPW+HLOs $GW_0$ gap, obtained in Sec. III.A, which is 6.27 eV in YAMBO code.}
\end{figure}
In Fig.~\ref{ZPR} we also show the $T$-dependence of the fundamental band gaps, which is similar to that
of the direct one.
Direct comparison of the $T$-dependent band gaps with the experiment is difficult due to the
lack of experiments.
Therefore, we evaluate the reliability of our results by considering similar systems.
In cubic BN, renormalization induces a 0.26 eV reduction of the quasiparticle band gaps in
an earlier theoretical study~\cite{Antonius2015}.
For monolayer BN, a similar theoretical result of 0.27 eV was reported~\cite{Mishra2019}.
The zero-point renormalization of the indirect optical gap of BN nanotubes is experimentally
estimated to be 0.225 eV~\cite{Vuong2018, Du2014}.
All these numbers are in close proximity to our results, when compared with the 0.4 eV
reduction of quasiparticle band gap in Ref.~\onlinecite{Hunt2020} and the 0.07 eV widening
of the optical band gap in Ref.~\onlinecite{Marini2008}.
In order to visualize the electron phonon coupling strength for a given state $|n{\mathbf{k}}$$>$, we compute the generalized Eliashberg function
\begin{equation}
\begin{split}
g^2F_{n\mathbf{k}}(\omega)=\frac{1}{N}\sum_{n^{'}\mathbf{q}\nu}\left[
\frac{|g_{nn^{'}\nu}^{\rm Fan}|^{2}}
{\epsilon_{n\mathbf{k}}-\epsilon_{n^{'}\mathbf{k}-\mathbf q}}-\frac{1}{2}
\frac{g_{nn^{'}\nu}^{\rm DW}}
{\epsilon_{n\mathbf{k}}-\epsilon_{n^{'}\mathbf{k}}}\right]
\\
\delta(\omega-\omega_{\mathbf{q}\nu}).
\end{split}
\label{Eliashberg}
\end{equation}
The band gap Eliashberg function is defined as $F_{\rm g}=F_{\rm c}-F_{\rm v}$, where the $\rm c$ $(\rm v)$ refers to a given conduction (valence) state.
Eliashberg functions are helpful to analyze contributions from different vibrational modes.
Combining Eqs.~(\ref{band}), (\ref{Fan}), (\ref{DW}) and (\ref{Eliashberg}), it is clear that the
$g^2F$ is directly proportional to band gap renormalization
\begin{equation}
\Delta E_{g}(T)\propto \int d\omega g^2F_{g}(\omega)[N_{\mathbf{q}\lambda}(T)+1/2].
\label{Eliashberg2}
\end{equation}
Fig.~\ref{Eliash}(a) shows the Eliashberg function at $H$ for $\rm v$ and $\rm c$ of the direct band
gap in Fig.~\ref{FIG1}.
At high frequencies, $F_{\text{g}}$ presents the most important peaks spanning
from 1400 to 1600 $\rm cm^{-1}$ owing to scattering events at $\rm v$.
Below the phonon dispersion gap, two dominant peaks exist between 600 and 850 $\rm cm^{-1}$.
Fig.~\ref{Eliash}(b) shows 12 phonon modes with three acoustic modes ZA, TA and LA, and nine
optical modes ZO, TO and LO.
The notation $Z$, $T$, $L$ represents out-of-plane modes, in-plane transverse modes and in-plane longitudinal modes respectively.
In Fig.~\ref{Eliash}(d) we project $g^2F$ to each mode.
To see more clearly, we separate it into two parts and only the representative modes are shown,
below 900 $\rm cm^{-1}$ in Fig.~\ref{Eliash}(e) and above that in Fig.~\ref{Eliash}(f).
In the low-frequency part, the electronic states couple mainly with the 3rd branch $\rm {LA}$ and
the 4th one $\rm {TO_1}$ around 600 $\rm cm^{-1}$, which reduce the band gap.
In the high-frequency part, the electronic states couple mainly with the 12th branch $\rm LO_{3}$
spanning from 1400 to 1600 $\rm cm^{-1}$, which is mainly responsible for the reduction of band gap with respect to the clamped crystal structure calculation.
From the analysis of the $T$-dependent generalized Eliashberg function, the acoustic modes are responsible for the decrease of the band gap as the $T$ increases (please see Fig. S3).
It is worth pointing out that contributions from the first two modes ZA and TA at around
350 $\rm cm^{-1}$ tend to increase the band gap.
However, this is negligible.
In Fig.~\ref{Eliash}(c), we also show results from the Eliashberg function at $K$ (close to $T_1$) for v
and at $M$ for c of fundamental band gap, of which the dominant
peaks are almost the same as those of the direct band gap.
\begin{figure}
\includegraphics[width=0.93\linewidth]{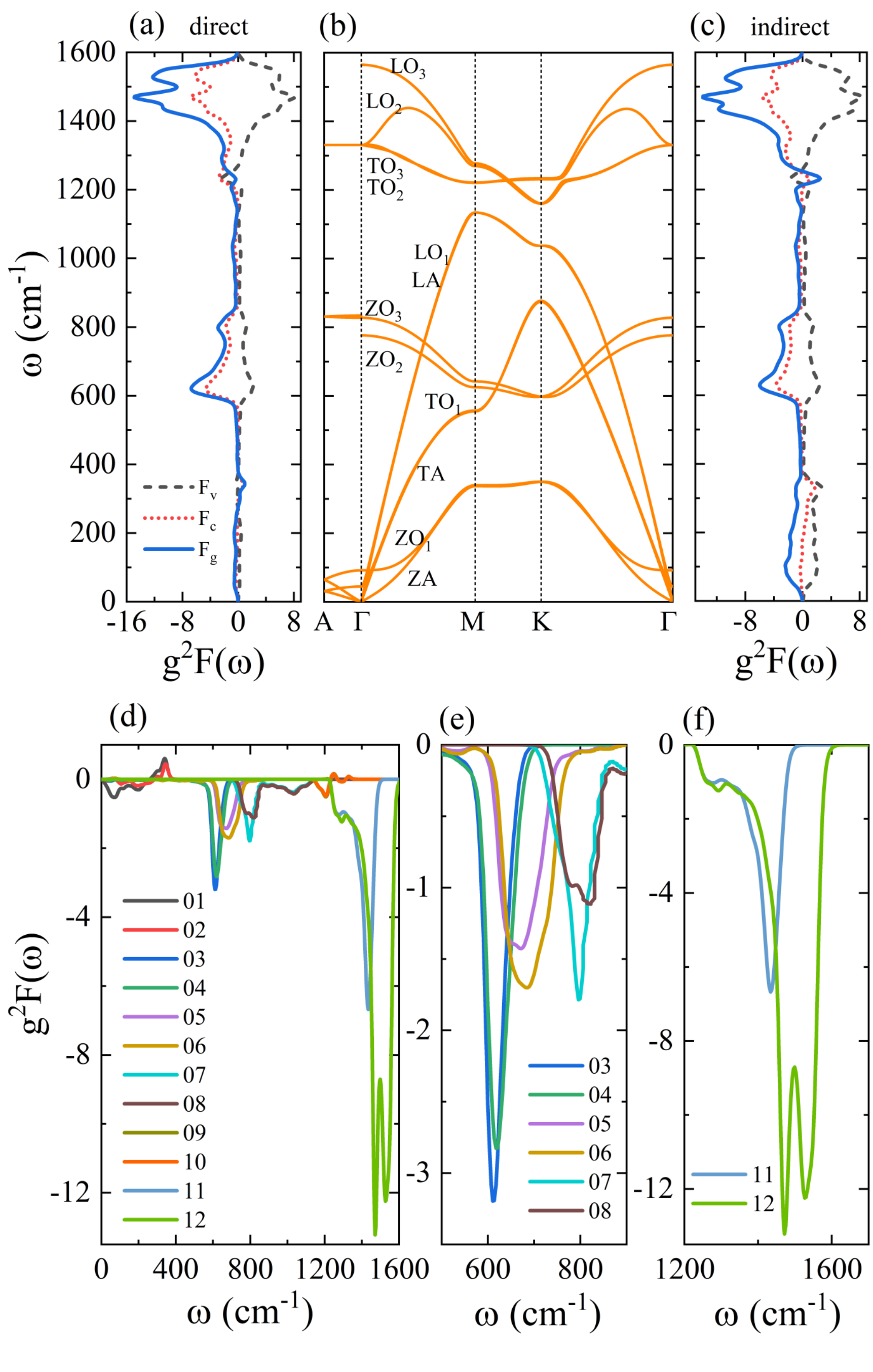}
\caption{\label{Eliash} (a) Generalized electron-phonon Eliashberg function for v (black dashed line), c (red dotted line), and band edge (blue solid line) of direct band gap. (b) Phonon dispersion along selected symmetry points. (c) Eliashberg function of fundamental indirect band gap. (d) The direct band-edge Eliashberg function projected on each phonon modes. Only the most representative phonon modes are considered (e) below 900 $\rm cm^{-1}$ and (f) above 1200 $\rm cm^{-1}$.}
\end{figure}

\subsection{Exciton and phonon coupling}
\begin{figure}
\includegraphics[width=0.90\linewidth]{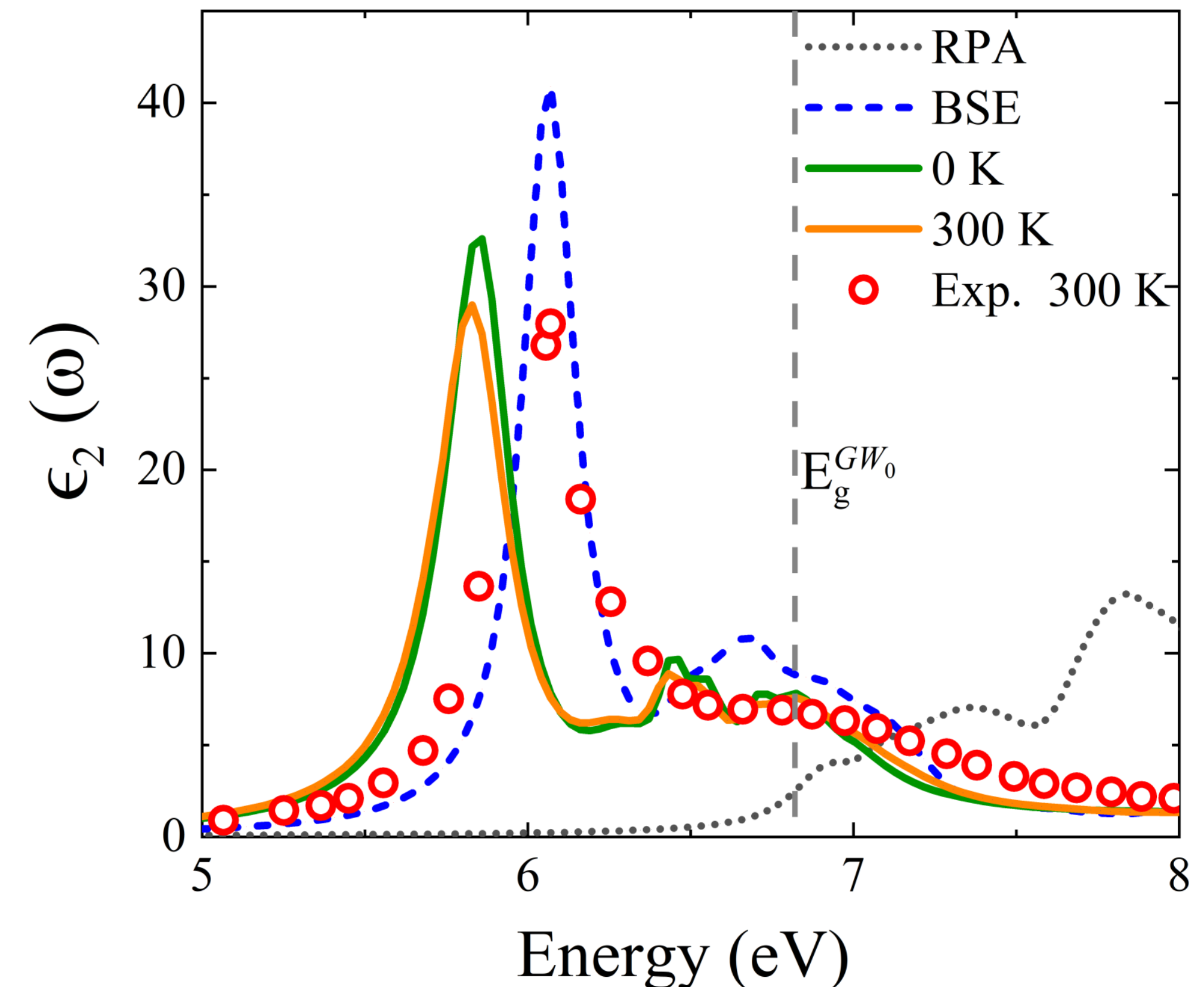}
\caption{\label{AHC} Comparison between measured and calculated optical absorption of $h$-BN: The experimental spectra~\cite{Tarrio1989} (red circles), calculation without EPIs at RPA (black dotted line) and BSE (blue dashed line) level, and calculation with EPIs at BSE level at 0 K (green solid line) and 300 K (orange solid line).}
\end{figure}

Now we calculate the optical spectra using the quasiparticle energies corrected by $GW_0$ with
HLOs, taking EPIs into account.
The electron-hole interactions are described using the BSE and the excitonic effects
are included in two steps.
With clamped crystal structure, the absorption spectra for $h$-BN with and without
electron-hole interactions are calculated first and shown in Fig.~\ref{AHC}.
With respect to the random-phase approximation (RPA) result, the excitonic effects substantially
redshift the optical spectrum peak and increase the intensity of absorption.
The BSE results with static nuclei gives the first bright bound exciton state at 6.06 eV, corresponding
to an exciton binding energy of 0.76 eV.
It is mainly contributed by the electron transition around $K$ point in the BZ.
More details on the excitonic effect are showed in Fig. S1, S2.
To consider the exciton-phonon couplings, we follow the treatment of Ref.~\onlinecite{Marini2008}.
In the BSE, the quasiparticle energies for the electron and hole states replaced by the
EPIs-renormalized ones, with static-nuclei BSE kernel kept.
In so doing, the $T$-dependent excitonic Hamiltonian is given by
\begin{equation}
\begin{split}
H_{\text{ee}^{'}\text{hh}^{'}}(T)
=[E_{\text{e}}+\Delta{E_{\text{e}}(T)}-E_{\text{h}}-\Delta{E_{\text{h}}(T)}]\delta_{{\text{eh}},{\text{e}}^{'}{\text{h}}^{'}} \\
+(f_{\text{e}}-f_{\text{h}})\Xi_{{\text{ee}}^{'}{\text{hh}}^{'}}.
\end{split}
\label{hamiltonian}
\end{equation}
Here, $E_{\text{e(h)}}$ and $f_{\text{e(h)}}$ are the quasiparticle energy of the electron (hole) and their
occupation number.
$\Delta{E_{\text{e(h)}}(T)}$ stands for the renormalization of electron and hole quasiparticle energies
after EPIs are included using the method discussed in Sec.~\ref{bandgaprenorm}.
$\Xi$ is the Bethe-Salpeter (BS) kernel, which are calculated using KS orbitals and energies.
Note that becuase $\Delta{E_{\text{e(h)}}(T)}$ are complex numbers, the excitonic Hamiltonian is non-Hermitian.
The dielectric function depends explicitly on $T$~\cite{Villegas2016}, through
\begin{equation}
\epsilon(\omega,T)\propto\sum_{X}\left|S_ {X}(T)\right|^2Im\left[\frac{1}{\omega-E_X(T)}\right],
\end{equation}
where $S_X(T)$ is the oscillator strength of each exciton and $E_X(T)$ is the eigenvalues of
Hamiltonian in Eq.~[\ref{hamiltonian}].
Fig.~\ref{AHC} shows that the main peak position slowly shifts to lower energies with increasing
$T$, and the linewidth broadens.
The main peak position of 0 K is at 5.86 eV, which shifts to 5.83 eV at 300 K.
Comparing with the results obtained using clamped crystal structure, this means that the
exciton-phonon couplings shift the absorption peak to lower energies by 0.21 eV at
0 K and by 0.24 eV at 300 K, which is in accordance with the extent of the band gap
renormalizations due to EPIs.
This is qualitatively different from the blueshift of 0.07 eV in Ref.~\onlinecite{Marini2008}.
The close proximity between our quasiparticle band gap renormalizations and optical band
gap renormalizations also means that the quasiparticle band gaps reductions are mainly
responsible for the redshift of the optical absorption peak.
To understand the discrepancy with Ref.~\onlinecite{Marini2008} in a clean manner, we first go
back to the optical absorption spectrum with clamped crystal structure.
The first optical absorption peak in Ref.~\onlinecite{Marini2008} is at 5.75~eV, while ours is
at 6.07~eV.
In Table~\ref{table1}, we can see a difference of 0.35 eV between our $GW_0$ direct quasiparticle
gap (6.81 eV) with HLOs and the QE+YAMBO value (6.46 eV) without especially constructing
channels for the high-lying electronic states in the PPs.
This difference of quasiparticle band gap is close to the redshift of 0.32 eV of their optical gap
with respect to ours.
Therefore, although the final optical spectra look similar, discrepancies do exist for both the quasiparticle band structures and the influence of exciton-phonon coupling.
Understanding these discrepancies is very important for a systematically improvable theoretical
description of the $h$-BN spectrum.
We believe the above analysis can give a clear picture on this.
Adding HLOs is very important in this sense, because otherwise the accuracy will become completely unacceptable ($\sim$0.5 eV) for state-of-the-art \textit{ab initio} theoretical methods.

\section{Conclusions}

Using a combination of \textit{ab initio} theoretical methods, we studied the
band diagram and optical absorption spectrum of hexagonal boron nitride ($h$-BN).
The focus is on understanding how the completeness of basis set for $GW$ calculations and how electron-phonon interactions (EPIs) impact on them.
The completeness of basis set is an issue which had been often overlooked in previous optical
spectra calculations of $h$-BN.
We found that it is crucial in providing converged quasiparticle band gaps.
By including HLOs in the all-electron linearized augmented plane
waves (LAPWs)-based $GW$ calculations, the quasiparticle direct and fundamental indirect
band gaps are widened by $\sim$0.2~eV, giving values of 6.81 eV and 6.25 eV respectively
at the LDA-based $GW_0$ level.
This proves the best starting point for later simulations of excitonic effects EPIs.
Upon including electron-phonon coupling, these gaps reduce to 6.62 eV and 6.03 eV respectively
at 0 K, and 6.60 eV and 5.98 eV respectively at 300 K.
Using clamped crystal structure, the first peak of the absorption spectra is at 6.07 eV, originating
from the direct exciton contributed by electron transitions around $K$ in the Brillouin zone (BZ).
After including the EPIs-renormalized quasiparticles in the BSE, the exciton-phonon coupling
shifts the first peak to 5.83 eV at 300 K
This is lower than the experimental value of $\sim$6.00 eV.
But the accuracy is acceptable to an \textit{ab initio} description of excited states, when no
fitting parameter are used.

\begin{acknowledgements}
The authors are supported by the National Basic Research Programs of China
under Grand Nos. 2016YFA0300900 and 2017YFA0205003, the National Science Foundation of China under
Grant Nos 11774003, 11934003, 11634001 and 21673005, the Special Fund for Strategic Pilot Technology of Chinese Academy of Sciences XDC01040000. The computational resources were supported by the
High-performance Computing Platform of Peking University, China.
The authors want to thank D. Nabok, E. Cannuccia, A. Marini, Q. J. Ye, X. F. Zhang and F. Liu
for many fruitful discussions.
\end{acknowledgements}

\end{document}